\documentclass[epj]{svjour}
\usepackage{graphicx}
\usepackage{txfonts}
\newcommand{\unipd}{1}
\newcommand{\infnpd}{2}
\newcommand{\infnlns}{3}
\newcommand{\unict}{4}
\newcommand{\infnlnl}{5}
\newcommand{\univen}{6}

\begin{document}

\title{A new study of  $^{10}$B(p,$\alpha$)$^{7}$Be reaction at low energies}
%Realisation of isotopic oxygen solid targets by tantalum anodic oxidation}

\author{
	A.\,Caciolli \inst{\unipd}\thanks{e-mail: caciolli@pd.infn.it} \and
	R.\,Depalo \inst{\unipd} \and
	C.\,Broggini \inst{\infnpd} \and
	M.\,La\,Cognata \inst{\infnlns} \and
	L.\,Lamia \inst{\unict} \and
	R.\,Menegazzo \inst{\infnpd} \and
	L.\,Mou \inst{\infnlnl} \and
	S.\,M.\,R.\,Puglia \inst{\infnlns} \and
	V.\,Rigato \inst{\infnlnl} \and
	S.\,Romano \inst{\infnlns,\unict} \and
	C.\,Rossi\,Alvarez \inst{\infnlnl} \and
	M.\,L.\,Sergi \inst{\infnlns} \and 
	C.\,Spitaleri \inst{\infnlns,\unict} \and
	A.\,Tumino \inst{\infnlns,\univen}\\
	}% author% etc
\institute{
	Dipartimento di Fisica e Astronomia, Universit\`a di Padova, and INFN, Sezione di Padova, Via Marzolo 8, Padova, Italy %9
	\and
	INFN, Sezione di Padova, via Marzolo 8, 35131 Padova, Italy % 1
	\and 
	INFN, Laboratori Nazionali del Sud, Catania, Italy	
	\and
	Dipartimento di Fisica e Astronomia, Universit\`a degli Studi di Catania, Catania, Italy
	\and
	INFN, Laboratori Nazionali di Legnaro, Legnaro, Italy % 12
	\and
	Facolt\`a di Ingegneria e Architettura, Universit\`a degli Studi di Enna ``Kore", Enna, Italy
	}% institute%

\date{\today}
\abstract{
The $^{10}$B(p,$\alpha$)$^{7}$Be reaction is of great interest since it has many applications in different fields of research such as nuclear astrophysics, nuclear physics,  and models of new reactors for clean energy generation. This reaction has been studied at the AN2000 accelerator of the INFN National Laboratories of Legnaro (LNL). The total cross section has been measured in a wide energy range (250 $-$ 1182 keV) by using the activation method. The decays of the $^7$Be nuclei produced by the reaction were measured at the  low counting facility of LNL by using two fully shielded high-purity germanium detectors. The present dataset shows a large discrepancy with respect to one of the previous data at the same energies and reduces the total uncertainty to the level of 6\%. An R-matrix calculation has been performed on the present data   using the parameters from previous Trojan Horse measurements for the 10 and 500 keV resonances. The present data do not lay on the R-matrix fit in one point suggesting the existence of a $^{11}$C level not  observed yet. Further nuclear investigations are needed to confirm this hypothesis.
 % end of PACS codes
} %end of abstract
% =========================================================
\authorrunning{}
\titlerunning{}
% =========================================================
\maketitle

\section{\label{sec:level1}Introduction}

The interest of the scientific community to study boron destroying reactions has  increased in the last years mainly because of their importance in several domains, from nuclear physics to plasma physics and astrophysics. 
%Among the light elements, boron is  largely used as semiconductor dopant or neutron absorber in many industrial applications and technological solutions \cite{lagoyannis2015}. 
Natural boron consists of two stable isotopes:  $^{11}$B (80.1\%) and  $^{10}$B (19.9\%), and nuclear processes involving both of them are almost equally important.

Referring to nuclear physics studies, the (p,$\alpha$) channel of boron induced reactions has been recently studied by \cite{stave2011,spraker2012} because of its role as possible source of aneutronic energy production, like in plasma induced fusion reactions. The possibility to reduce the presence of neutrons in the exit channel, offered by the reaction on $^{11}$B, promotes this solution with respect to the deuteron-tritium reaction where a large emerging neutron flux is estimated \cite{labaune2013}. 
Besides the role played by the $^{11}$B(p,$\alpha$)$^8$Be reaction \cite{stave2011,spraker2012},  the presence of the $^{10}$B isotope in the nuclear fuel could also trigger the $^{10}$B(p,$\alpha$)$^7$Be reaction in its two channels, namely, the $^{10}$B(p,$\alpha_0$)$^7$Be and the $^{10}$B(p,$\alpha_1$)$^7$Be$^*$, leading to the production of $^7$Be in its ground and first excited states, although the latter is strongly suppressed \cite{Rauscher1996,Brown1951}.    
The production of  radioactive $^7$Be  could also pose radiation-safety problems \cite{lagoyannis2015,kafkarkou2013} because of its relatively long half-life, T$_{1/2}$=53.22$\pm$0.06 d \cite{Tilley2002}.
A new  precise cross section measurements of the $^{10}$B(p,$\alpha$)$^7$Be reaction could also be used to understand its impact in future fusion reactor projects.
 
With regards to nuclear structure, proton induced reactions on $^{10}$B could provide useful information on the structure of $^{11}$C. The level scheme of $^{11}$C has recently been  investigated in Ref. \cite{yamaguchi,Freer2012} by means of the inverse process $^7$Be+$\alpha$, used to measure the proton or alpha partial widths of the $^{11}$C excited states. These results have been compared with the ones extracted for the mirror $^{11}$B nucleus. The $^{10}$B(p,$\alpha$)$^7$Be reaction forms $^{11}$C as compound nucleus and  could be used to better understand its level scheme. 

Boron plays an important role in astrophysics and, together with the other light-elements lithium and beryllium, is a probe of stellar structure for both pre-main  sequence (pre-MS) \cite{tognelli11} or main-sequence (MS) stars \cite{boesg}. In more details, lithium, beryllium and boron start to be destroyed by (p,$\alpha$) reactions at different temperatures,  from $\sim$2.5$\times$10$^6$ K up to $\sim$4.5$\times$10$^6$ K, corresponding  to different depths of stellar interior \cite{boes2004}. Thus, as addressed in several works \cite{boes2004,boes76,dely2000,Somers2015}, their residual abundance can be used to trace mixing phenomena in stars. Additionally, in order to constrain stellar theoretical models, nuclear reaction cross sections need to be precisely measured at the corresponding Gamow energy. For such a reason, the impact of (p,$\alpha$) reactions induced on lithium, beryllium and boron have been recently discussed in \cite{Somers2015,Tognelli2012,Lamia2013,Somers2014,Tognelli2015,Lamia2015}.
%articles such as Tognelli et al. 2012 \cite{Tognelli2012}, Lamia et al. 2013 \cite{Lamia2013}, Somers et al. 2014 \cite{Somers2014}, Somers et al. 2015 \cite{Somers2015},  Tognelli et al. 2015 \cite{Tognelli2015}, Lamia et al. 2015 \cite{Lamia2015}. 
Focusing on boron isotopes, $^{11}$B and $^{10}$B are important not only for their absolute abundances but also for their relative abundances in pre-MS stars. In particular, the temporal evolution of the relative N($^{11}$B)/N($^{10}$B) abundance for different stellar masses and metallicities have been recently studied \cite{Lamia2015}, 
%The initial values of N($^{11}$B)/N($^{10}$B) were taken from the article of Prantzos et al. 2012 in which the authors give the value of N($^{11}$B)/N($^{10}$B)=4 for [Fe/H]=0 and N($^{11}$B)/N($^{10}$B)=4.5 for [Fe/H]=-1.5. 
showing an increase with time for stellar masses in the range 0.1$<$M/M$_\odot <$0.3. This is expected, because the depletion of $^{10}$B and $^{11}$B occurs at different ages, due to the slightly different ignition temperatures (about 4$\cdot$10$^6$ K for $^{10}$B and 5$\cdot$10$^6$ K for $^{11}$B), with the $^{10}$B burning starting at younger ages. Consequently, the N($^{11}$B)/N($^{10}$B)  value is expected to increase with time. \\
For such a reason, both isotopes are important in the general picture of pre-Main Sequence stars and here we are going to focus our interest on the $^{10}$B(p,$\alpha$)$^7$Be reaction channel.
%. These elements are destroyed at different depths in  stellar interiors and  residual (atmospheric) abundances can be used  to constrain mixing phenomena occurring in such stars \cite{boesg}. Boron burning is triggered at temperatures T$\geq$5$\cdot$10$^6$ K  and takes place mainly through (p,$\alpha$) processes, with a Gamow peak \cite{Rolfs1988} centered at about 10 keV.
In this context, the $^{10}$B(p,$\alpha_0$)$^7$Be reaction, with $^7$Be left in its ground-state, is of particular interest. Its cross section  at the  Gamow energy (E$_G$) is dominated by the contribution of the 8.699 MeV  (J$^\pi$=$\frac{5}{2}^+)$ $^{11}$C state, corresponding to an s-wave resonance centred at about 10 keV.\\

Most of the $^{10}$B(p,$\alpha_0$)$^7$Be cross section measurements are reported in the widely adopted NACRE compilation \cite{Angulo1999}  and in Refs.\cite{Brown1951,lagoyannis2015,kafkarkou2013,Cronin1956,Youn1991,Szabo1972,Wiescher1983,Angulo1993,spitaleri2014,Lombardo2016,atdata79}, referring to different experiments and ranging from 5 keV up to more than 6 MeV.

The differential cross section measurements of \cite{kafkarkou2013} provide a total $^{10}$B(p,$\alpha_{0,1}$)$^7$Be cross section, which is on average 16\% higher than previous experiments in the energy range between 2 and 6~MeV.
At about 1 MeV, the experimental data of \cite{atdata79} refer to a thick-target activation measurement  for which a deconvolution procedure was operated  to derive the astrophysical S-factor when included in the NACRE compilation  \cite{Angulo1999}. In 1951, Brown et al. measured the cross section for both the (p,$\alpha_0$) and (p,$\alpha_1$) channels, finding that the $^{10}$B(p,$\alpha_1$)$^7$Be$^*$ branching is highly improbable  at energies below 1.3 MeV \cite{Brown1951}. After that, in 1956, Cronin \cite{Cronin1956} measured the angular distribution for both the two channels at energies above 1~MeV.
At low energies, i.e. E$<$100 keV,  data in Ref. \cite{Angulo1993} show an enhancement of  $S(E)$. This enhancement is  produced by the presence of a resonance at $E_{CM} \sim$ 10 keV  and by the electron screening effect \cite{Assenbaum1987,Strieder2001}, thus making  extrapolation procedures the only tool to separate these effects and evaluate the $^{10}$B(p,$\alpha_0$)$^7$Be cross section around the 10 keV resonance.

It is worth to mention that in Ref. \cite{Angulo1993} the experimental data of Ref. \cite{Youn1991}, between 200 keV and 500 keV, have been scaled by a factor of 1.83 to obtain a better agreement of different datasets over a larger energy range. However, a detailed explanation of the discrepancy among different datasets is not discussed by \cite{Angulo1993}. 
A new experiment on the (p,$\alpha_0$) channel has been published recently \cite{Lombardo2016} in the 0.630-1.028 MeV energy range reporting the possible existence of two new $^{11}$C levels, but also demanding for new measurements in a wider energy domain.

To overcome the difficulties related to the electron screening effect and the suppression of the cross section at ultra-low energies, recent measurements of  $^{10}$B(p,$\alpha_0$)$^7$Be with the Trojan Horse Method (THM) \cite{spitaleri2014} have provided the bare-nucleus $S(E)$ in correspondence of the $\sim$10 keV resonance without the needs of extrapolation procedures, leading to the value of $S$(10 keV)=(3.1$\pm$0.6)$\cdot$10$^3$ MeV$\cdot$b at the top of the resonance.

To solve the above mentioned mismatch between the data of Youn et al. \cite{Youn1991} and the ones of Angulo et al. \cite{Angulo1993}, a new experimental study in the region 200$<E_{cm}<$1000 keV is necessary. These new precise data could also provide to THM measurements a more extended energy region for normalisation purpose. 

In view of these points, a new measurement of the $^{10}$B(p,$\alpha$)$^7$Be cross section has been performed in the 200$<E_{cm}<$1000 keV energy region by means of the activation  technique. A detailed discussion of the experimental setup and data reduction is given, together with the comparison between the measured experimental data and R-matrix calculations.

\section{Experimental Setup}
\label{sec:experiment}

The setup was installed at the AN2000 Van De Graff accelerator of the INFN National Laboratories of Legnaro (LNL), which is able to provide a  proton beam  with beam intensity of 200-300~nA. 
The beam energy was calibrated to a precision of 1~keV using the 992 and 632~keV narrow resonances of the $^{27}$Al(p,$\gamma$)$^{28}$Si reaction \cite{26Al}. 
The calibration was verified at 2~MeV with Rutherford Backscattering Spectrometry (RBS) measurements on SiO$_2$ samples.

\begin{figure}[htb]
\centerline{%
\includegraphics[width=\columnwidth]{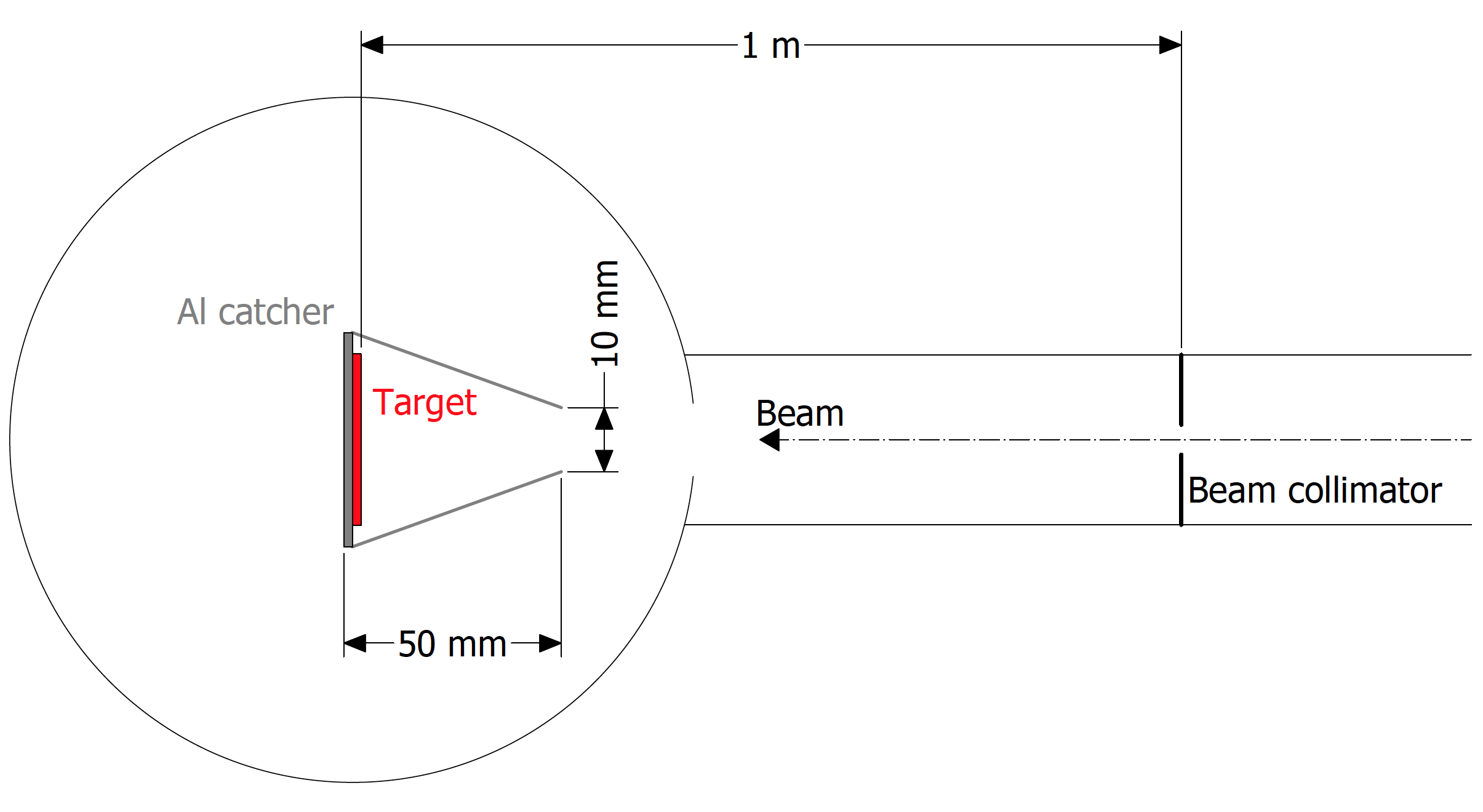}}
 \caption{A scheme of the experimental setup, see text for details. The Al catcher dimension are reported in the figure.} \label{fig:setup}
 \end{figure}

The proton beam entered  the scattering chamber passing through a collimator of 9 mm diameter.  
This is the only aperture present in the scattering chamber. The beam charge is collected on the whole chamber that works  as Faraday cup.
The beam size is defined by collimation slits placed at 1~m distance from the target. 
A sketch of the experimental setup is shown in figure \ref{fig:setup}.
%===========================
%===========================

Targets were made of  boron powder enriched up to 93\% in $^{10}$B evaporated on thin carbon layers ($\sim$20 $\mu$g/cm$^2$), and they were installed at the center of the chamber at 90$^\circ$ with respect to the incoming proton beam. 
Targets were made with three different thicknesses: 25, 60, and 100 $\mu$g/cm$^2$. 
Targets with the same thickness were evaporated at the same time in order to ensure their uniformity and two samples of each group were not irradiated and used to study their properties as discussed in section \ref{sec:analysis}.

The $^{10}$B(p,$\alpha$)$^{7}$Be cross section has been measured through activation technique \cite{Iliadis_book}.

Since the kinetic energy of emitted $^7$Be ions is enough to escape the boron targets\footnote{The energies of backward emitted $^7$Be are in the range of 340$-$770~keV depending on the angle and on the proton beam energy ($E_p$) and they were calculated using LISE++  \cite{LISE}}, 0.2 mm thick Al catchers were positioned all around the target. A foil was placed behind the target to catch the $^{7}$Be ions forward emitted, while a cone-shaped catcher was mounted in front of the target. The cone was 50 mm long and has a hole of 10 mm diameter to allow the entrance of the beam. 
%The target was shifted forward with respect to the cone base to avoid any possible $^7$Be nuclei escaping at 90$^\circ$.
Using this geometry the catchers cover the 99.8\% of the total solid angle. Moreover, their thickness is sufficient to stop the $^7$Be ions produced by the reaction at all investigated beam energies.
%an aluminium cone was placed in front of the target.
%Similarly an Al foil  was also placed behind the target to catch $^7$Be ions forward emitted.
%The sheets used for the cone and the foil were 0.2 mm thick,
%that is sufficient to stop the $^7$Be ions produced by the reaction at all investigated beam energies.
%The target was shifted forward with respect to the cone base to avoid any possible $^7$Be nuclei escaping at 90$^\circ$.
%The cone was 50 mm long and it had an hole of 10 mm diameter to allow the entrance of the beam. This way the Al catchers cover the 99.8\% of the total solid angle.
The size of the beam spot was checked on graph paper before any irradiation  and it was kept at 3~mm $\times$ 3~mm maximum to guarantee that the beam did not hit the cone before reaching the target.

Targets were irradiated at each energy for several hours depending on the beam energy and expected cross section. 
Once the irradiation was completed, both the  target and the Al catchers were  inserted in the low background $\gamma$-counting facility of the LNL \cite{Xhixha}, thus allowing for the detection of the produced $^7$Be nuclei.
$^7$Be decays with (10.44$\pm$0.04)\% probability to the first excited state of $^7$Li. The de-excitation of $^7$Li to the ground state produces a 478 keV $\gamma$ ray that can be used to monitor the $^7$Be decay.

The setup is composed by two high purity germanium detectors of 80\% relative efficiency facing each other. The samples were arranged to be  coaxial with the  two detectors at a distance of 2.5 cm  from each surface.
The entire setup is shielded by few cm of copper and 10~cm of lead  to reduce the environmental background by about 2 orders of magnitude for $\gamma$ energies below 3~MeV.
The $\gamma$-detection efficiency  for the 478~keV $\gamma$ line was measured by using a  $^7$Be source, calibrated with a precision of 2\%. 
An additional 2\% uncertainty was added due to a possible misplacement of the source from the center of the two HPGe. The final value for the absolute full-energy peak efficiency is 0.081$\pm$0.002.

\section{Analysis and Results}\label{sec:analysis}

Targets and catchers were measured separately. 
Since  $^7$Be has a relatively long half-life, all samples were analysed several times in the three months following the completion of the experiment to monitor the $^{7}$Be decay in a period longer than the $^{7}$Be half-life.

\begin{figure}[htb]
\centerline{%
\includegraphics[width=\columnwidth]{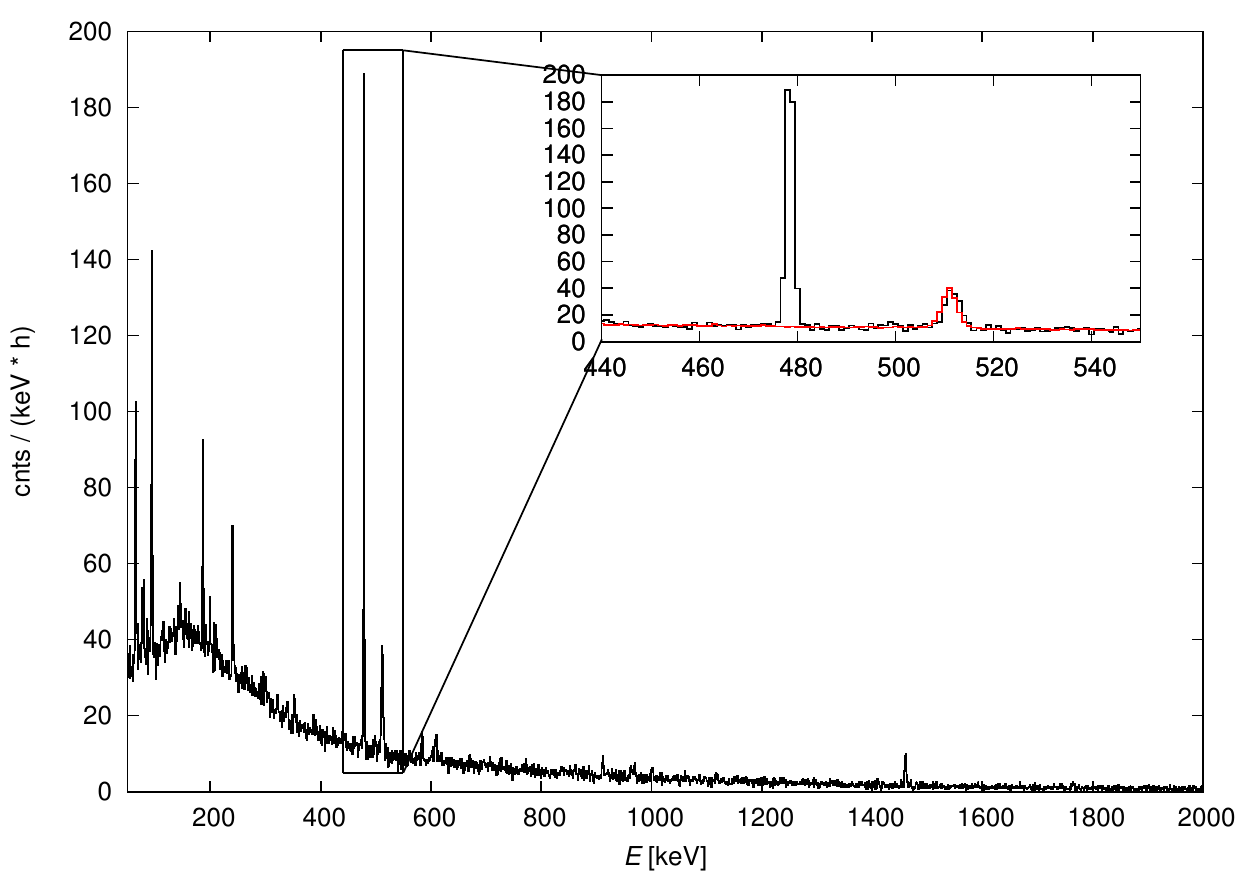}}
 \caption{The sum of the spectra obtained with the two germanium detectors in 6 hours. A total charge of  5.2~mC was accumulated at $E_p$ = 279 keV on 60~$\mu$g/cm$^2$ thick target, and the counting was started $\sim$12~h after the irradiation. A zoom of the region of interest for the 478~keV peak of the $^7$Be decay is  shown where the 511 $\gamma$ line is also visible. In the inset a 100 h background spectrum is also shown in red for comparison. No structures at 478~keV are present in the background.} \label{fig:spectrum2HPGe}
 \end{figure}

A typical spectrum, obtained after 6 hours counting  of a target irradiated at the lowest proton energy ($E_p$ = 279~keV), is shown in Figure \ref{fig:spectrum2HPGe}.
%===========================
%===========================

The number of $^7$Be nuclei produced in each irradiation was deduced from the analysis of the activation spectra. 
In particular, for each sample the initial number of $^7$Be nuclei was deduced using the exponential decay formula for each acquired spectrum and then those results were compared, finding an optimal agreement within 1 standard deviation. Then, the weighted average was adopted to obtain the final results and the statistical uncertainty.
%The analysis of spectra, taken at different time after the irradiation, show a good agreement in the number of $^7$Be nuclei 
%Since measurements taken at different times were in agreement with each other within 1 standard deviation,  the weighted average was adopted to obtain the final results and the statistical uncertainty. 
Targets and cones samples were analysed separately and then  summed to obtain the total amount of $^{7}$Be produced at each irradiation energy: $N_{^7Be}$. 

$S(E)$  was obtained  using the following formula:
%%==========================================
\begin{equation}\label{eq1}
S(E) = \frac{N_{^7Be}}{N_p}\frac{1}{\int_{E_p - \Delta E}^{E_p} \frac{(m_t + m_p)\exp^{-4.734\sqrt{\frac{m_t + m_p}{Em_t}}}}{E\cdot m_t \cdot\epsilon (E)}\mbox{d}E},
\end{equation}
%%==========================================
where $N_p$ is the number of impinging protons  during the irradiation, $m_p$ and $m_t$ the proton and $^{10}$B masses, and $\epsilon$ is the effective stopping power \cite{Iliadis_book}. The integral is done along the beam energy loss $\Delta E$ in the target. 
In eq. \ref{eq1} we assumed that  $S(E)$ is constant over $\Delta E$ and that the $^7$Be decays during the irradiation are negligible. We implemented also an $S(E)$ curve based on the previous data in literature or a recursive approach for the $S(E)$ behaviour without finding differences in the final S-factor results.
The effective energy of each irradiation was obtained calculating the average energy weighted with the exponential trend of the cross section \cite{Iliadis_book}.

The results are reported in table \ref{tab:results}, while systematic errors in Table \ref{tab:systerrors}.  They were combined in quadrature to obtain the total error. 

% and shown in figure \ref{fig:results} with previous  data in literature. 
%==========================================
\begin{table}
\centering
\caption{The astrophysical S-factor values obtained in the present work. The errors reported are only statistical.}\label{tab:results}
\begin{tabular}[!h]{ccc}
\hline \hline
$E_{CM}$[keV]	&	$S$[MeV b]		&	$\Delta S_{stat}$[MeV b]	\\
\hline
249	&	19.96	&	0.16					\\
347	&	14.47	&	0.06					\\
446	&	12.72	&	0.07					\\
548	&	9.34		&	0.13					\\
647	&	13.33	&	0.07					\\
749	&	10.75	&	0.08					\\
900	&	16.61	&	0.13					\\
1182	&	20.77	&	0.09					\\
\hline												
\hline																			
\end{tabular}
\end{table}
%==========================================	

For each type of targets two samples were not irradiated and analysed to determine their characteristics.
Targets thickness and elemental composition (including oxygen and nitrogen contaminants) were measured by using  Rutherford backscattering and nuclear reaction analysis  at the AN2000 accelerator.
The analysis of the two samples gave results in very good agreement for all groups and the maximum difference in composition was taken as size of the uncertainty (see table \ref{tab:systerrors}). 
Stopping powers are taken from SRIM \cite{SRIM}. The stopping power  was varied during the analysis within its uncertainty and the effect on the final astrophysical S-factor  was adopted as systematic error (0.5\%) and summed up with the error due to the target characterisation in Table \ref{tab:systerrors}.

The number of $^7$Be ions escaping from the the cone hole is negligible (0.2\%) assuming an isotropic angular distribution in the center of mass system. 
Since in the whole energy range of the present experiment no data on angular distributions are reported, we estimated the uncertainty due to this assumption testing 3 different angular distributions: the cos($\theta$) function, the one reported by Youn \emph{et al.} at 395 keV \cite{Youn1991}, and the one reported by Cronin at 1 MeV \cite{Cronin1956}  (to compare with experimental data at the upper and lower bounds of our measured energies).
We obtain the maximum deviation from an isotropic behavior for a $\cos(\theta)$ distribution and we assumed this value as a systematic uncertainty.
The effect of possible backscattering of $^7$Be ions on the aluminium cone has been evaluated with the TRIM code. Simulations were done assuming $^7$Be ions  impinging on Al layer with a maximum angle of 60$^\circ$ with respect to the perpendicular. In all simulations, less than 1\% of the total  $^7$Be was  backscattered by the Al.

%==========================================
\begin{table}
\centering
\caption{Systematical error budget.}\label{tab:systerrors}
\begin{tabular}[!h]{lc}
\hline \hline
Source	&	Error [\%]		\\
\hline
Charge	&	1					\\
Beam Energy	&	0.1$-$1.4					\\
Target	&	4				\\
HPGe efficiency	&	2.9					\\
Ang. Dist.	&	2					\\
$^7$Be backscatter 	&	1					\\
\hline
Total	&	6	\\
\hline												
\hline																			
\end{tabular}
\end{table}
%==========================================

\section{Discussion and Conclusions}

\begin{figure*}[htb]
\centerline{%
\includegraphics[width=0.8\textwidth]{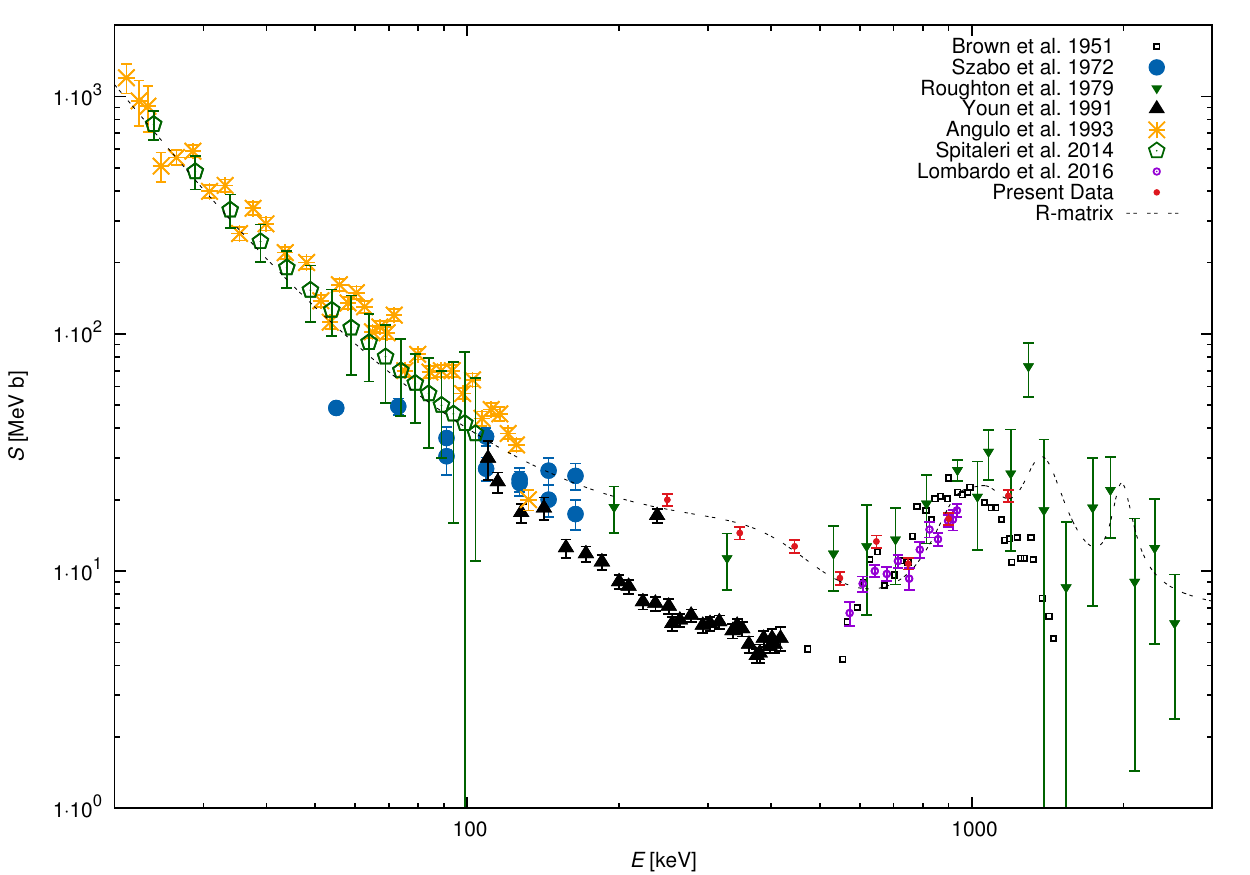}}
 \caption{The present S-factor data are compared with the existing data in literature. In figure the R-matrix fit is also reported as black line. See the text for details.} \label{fig:results}
 \end{figure*}

The $^{10}$B(p,$\alpha$)$^{7}$Be cross section has been measured in the energy range from 249 up to 1182 keV in the center of mass system, spanning a cross section range from 6 mb up to 229 mb. The data are shown in figure \ref{fig:results} together with the previous data in literature. In figure \ref{fig:results} statistical and systematic errors are summed together. 
The new data cover an energy range poorly explored in previous works.
In particular, the present data show a  discrepancy with respect to Youn \emph{et al.} \cite{Youn1991} by about a factor of 2. This value is similar to the one used by Angulo \emph{et al.} \cite{Angulo1993} to normalise the Youn \emph{et al.} data to their results, based on a very small overlapping energy range. 
The data in  \cite{Brown1951} and \cite{Lombardo2016} are in very good agreement with the present data. Only the two points at lowest measured energies in \cite{Brown1951}  show a lower S-factor with respect to the present one, but at these energies (below 1 MeV) the authors of \cite{Brown1951} state that the background due to scattered proton and the huge target thicknesses introduced several complications in the data analysis.
%In addition, the present data show also a different S-factor shape with respect to the Youn data. This affects extrapolations and normalisations based on the data in this energy range.
The agreement with the Roughton \emph{et al.} \cite{atdata79} data is very good, but in the present data the uncertainties are much smaller.
%===========================
%===========================

A simplified R-matrix analysis of the present-work data was performed, including the p and $\alpha$ channels only. 
In the calculation, the properties of the $^{11}$C ($S_p$ = 8.69 MeV) levels at 8.7 ($E_R$ = 10 keV) and 9.2~MeV ($E_R$ = 500 keV), leading to the $^{10}$B(p,$\alpha$)$^{7}$Be resonances  at 10 and 500~keV, were fixed to the values obtained in \cite{spitaleri2014}. 
Four more resonances were included in the calculation, namely, the $^{11}$C levels at 9.64~MeV (J$^\pi$=3/2$^-$, $E_R$ = 960 keV), 9.78~MeV (J$^\pi$=5/2$^-$, $E_R$ = 1.1 MeV), 10.08~MeV (J$^\pi$=7/2$^+$, $E_R$ = 1.4 MeV), 10.68~MeV (J$^\pi$=9/2$^+$, $E_R$ = 1.99 MeV). Resonance energies were fixed as reported in table \ref{tab:r-matrix}. All  these levels have been found to alpha decay with the exception of the 9.78~MeV. 
The total widths of these levels were fixed to the values in the literature \cite{Tilley2002}, while p and $\alpha$ widths were allowed to vary to reproduce the experimental data. 
Physical resonance parameters are given in Table \ref{tab:r-matrix}. 
%==========================================
\begin{table}
\centering
\caption{Resonance parameters used for the R-matrix fit.}\label{tab:r-matrix}
\begin{tabular}[!h]{cccccc}
\hline \hline
$E_X$ 		 &	$E_R$ & J$^\pi$	&	$\Gamma_p$ & $\Gamma_\alpha$ & $\Gamma_{tot}$ \\
 $[$MeV]		 &	 [MeV] & 	&	[MeV] & [MeV] & [MeV] \\
\hline
8.699 &	 0.01 &	5/2$^+$			& &	0.015 \\
9.2	& 0.50 &	5/2$^+$		& 0.0018	& 0.501	& 0.503\\
9.645	& 0.96 &	3/2$^-$	&0.031	&0.222	&0.252\\
9.78$^a$	& 1.09 &	5/2$^-$	&0.018	&0.221	&0.239 \\    
9.97$^{a,b}$	& 1.28 &	7/2$^-$		&	&	&	\\
10.083		& 1.39 &	7/2$^+$		&0.187	&0.047	&0.234\\
10.679		& 1.99 &	9/2$^+$		&0.106	&0.098	&0.204\\
\hline												
\hline
\multicolumn{6}{l}{$^a$ The $\alpha$ decay of this level is not reported in literature.}\\																			
\multicolumn{6}{l}{$^b$ This level is not used in the R-matrix fit.}\\																			
\end{tabular}
\end{table}
%==========================================
The introduction of the 9.78~MeV state and of a non-resonant contribution equal to 6~MeV~b (not included in \cite{spitaleri2014}) and constant over the whole considered energy range have proved necessary to reach a satisfactory matching with data ($\chi ^2$=11 for 7 points). 
 This way the huge deep in the S-factor curve shown in  \cite{spitaleri2014} is now removed.
The good quality of the R-matrix analysis is demonstrated in Fig. \ref{fig:results}, where the R-matrix calculation is shown as a black solid line.
One experimental point, at  $E$ = 647~keV, significantly deviates from the calculation, much more than the total uncertainty, and might point to the existence of another, maybe narrow, resonance still unknown. 
A compatible enhancement of the cross section at the same energy is also present in \cite{Brown1951}, but the authors do not discuss that feature, probably due to the experimental complications at these low energies. In addition, \cite{Lombardo2016} introduce a new level  at this energy in their R-Matrix analysis, whereas their signature seems to be less evident.
Further investigations with a smaller energy step would be necessary to clarify the point. 

An important feature, which these data confirm, is the interference pattern between the 10~keV and the 500~keV resonance, which is destructive above 500~keV. 
This is clearly shown by the decrease of the astrophysical factor above 500~keV. 
Though the present work allows us to constrain the contribution of the 500~keV resonance in the R-matrix,  the constructive interference with the 10~keV resonance is not able to reproduce the experimental data of \cite{Angulo1993} at about 100~keV. At lower energies these data lay above the curve, but they are in agreement with the Trojan Horse data within the errorbars. 
Finally, this work puts the THM result \cite{spitaleri2014} on sounder grounds since the THM data and the experimental points discussed here can be fit with the same R-matrix. It has to be noted that in the present R-matrix a non-resonant contribution has been introduced. Its contribution is negligible at energies lower than about 100~keV, while it is important above 500~keV.

A more detailed R-matrix calculation including gamma channels and differential cross sections is ongoing, though the main features have been addressed in this calculation.

\section*{Acknowledgments}

The authors are indebted to Leonardo La Torre for the help in setup installation and for running the  accelerator. We thank Massimo Loriggiola for the target production and the  LNL mechanical workshops   for technical support. Financial support by INFN and University of Padua (Grant GRIC1317UT) is gratefully acknowledged. This work has been partially supported by the Italian Ministry of the University under Grant RBFR082838 and ``LNS-Astrofisica Nucleare (fondi premiali)".



% \bibliographystyle{elsarticle-num} 
% \bibliographystyle{epjc} 
  %\bibliography{Bibliografia_Antonio.bib}

%%%% else use the following coding to input the bibitems directly in the
%%%% TeX file.
%%

\end{document}